\documentclass[12pt]{article}
\pdfoutput=1
\usepackage{graphicx}
\usepackage[style=nature]{biblatex}
\title{Encoding Multifunctional Nanostructured Metasurfaces}
\author
{Bo Xiao,$^{\ast}$ , Christian G. Carvajal, Sangeeta Rout
\\
\normalsize{Center for Materials Research, Norfolk State University,}\\
\normalsize{700 Park Ave, Norfolk, Virginia 23504, USA}\\
\\
\normalsize{$^\ast$To whom correspondence should be addressed. E-mail: bxiao@nsu.edu.}
}
\date{}

\begin{document}

\maketitle

Metasurfaces offer unconventional control of light to shape optical wavefronts within two-dimensional nanoscale structures. A nanostructured metallic thin film can establish an exclusive relationship between its surface structure and optical spectrum. The unique structural features, continuous metal coverage and smooth surface profile, can help obtain highly coherent plasmonic resonances that strongly depend on the structural parameters of nanostructured surfaces. It allows us to encode optical responses and information into a structural format on a variety of materials. We show experimentally that the nanostructured metasurfaces can be encoded to detect and display chemical and biomolecular interactions, show colorful graphic images, and create multichannel holograms.

\paragraph*{}
The advent of nanotechnology enables a major paradigm shift from conventional bulky optical components to nanoscale structures to control and manipulate intricate light-matter interactions. Metasurfaces as a new type of artificial material structures are capable of gaining unprecedented control of electromagnetic waves that are unattainable in natural materials \cite{Yu2014,Kildishev2013,Meinzer2014}. In the past decade, fundamental research blended with a focus on applications has experienced an explosive growth and already led to some fascinating developments such as metalenses, invisible cloaking, nanoantenna metasurfaces, hyperbolic metasurfaces \cite{Khorasaninejad2016,Ni2015,Yu2011,High2015,Ni2012}. In practice, such metasurfaces are typically made of two-dimensional (2D) nanostructures on the surfaces of certain supporting materials, which are usually not the same as the nanostructures. The optical responses through a metasurface depend on not only the structural parameters but the surrounding environment of the structure. Optical properties of nanostructures and their surroundings cannot be simply inferred from their bulk materials. Consequently, a realized metasurface design is not readily applied to different materials without modifying its design or adjusting its optical configuration. Here, we suggest a nanostructured metasurface and some general encoding guidelines that establish an exclusive and predefined relationship between structure and spectrum. Optical response, information or manipulation can be encoded as a structural format onto a variety of materials, which are of considerable interest in multifunctional applications. 

\paragraph*{}
Our idea for encoding multifunctional metasurfaces is based on a simple nanostructure platform sketched in Fig. 1A. In general, the platform is a surface relief, which has nanoscale elevations or depressions on a flat surface (Fig. 1A). The unique feature of the structure is its smooth profile (Fig. 1B) that is introduced intentionally in the fabrication process (Fig. S1). The whole surface relief is covered by a continuous metal thin film. Metals such as Ag or Au are used as the coating to form a plasmonic metasurface, which can excite surface plasmon polaritons at a certain frequency by proper designing nanostructure patterns. The smooth profile is a critical part of our design, which not only alleviates structural discrepancies and thus improves coherent plasmonic resonances but is compatible with many high-throughput nanofabrication methods \cite{Xiao2016,Nagpal2009b}. Most importantly, the smooth profile and continuous metal coverage together are able to distinguish plasmonic responses between the two interfaces: metal and supporting material, and metal and air. Fig. 1E shows photographs of the metasurface with elevation relief structures under normal illumination from laser diode sources, which were captured by a regular digital camera without an infrared filter. Each square represents one array in which nanostructures are arranged with a periodic spacing ranging from 400 to 899 nm (Fig. 1D). The brightness of every square indicates its transmission efficiency. The strongest transmission occurs on the nanostructure arrays where the periodic nanostructures at the interface of the metal and air satisfy the plasmonic momentum matching condition and excite coherent resonances \cite{Xiao2015}. Meanwhile, the dark squares show an opposite effect which is dominated by the plasmonic resonances in the interface of the metal and supporting material as well as the metal thickness. The metasurface with depression relief is more complicated under normal transmission where the supporting material exerts a strong influence on plasmonic resonances (Fig. S2). 

These metasurfaces can achieve similar spectral manipulation in a grazing incidence mode (Fig. 1F). Periodic nanostructures in one dimension can be treated as a diffraction grating, which satisfies the general grating equation: $m\lambda = d(sin\theta_{i}+sin\theta_{r})$, where m is the diffraction order and a light beam with a wavelength of $\lambda$ is incident on a diffraction grating with a periodic spacing of \textit{d} at an angle $\theta_{i}$ and diffracted at $\theta_{r}$. Consider a plane wave incident from a small angle ($sin\theta_{i}\approx1$). A constructive interference occurs at the surface normal ($sin\theta_{i}=0$) where $d\approx\lambda$ (Fig. 1C). Thus even a non-plasmonic grating can yield constructive interference from an incident monochromatic light beam as shown in Fig. 1E, where the nanostructured surface was formed by a polymer layer without a metal coating on a glass substrate. However, the general grating equation is unable to describe the plasmonic effect for the surfaces with a plasmonic thin layer. A metasurface with a 40 nm Ag thin film exhibit enhanced and diminished reflections for the nanostructure arrays at the places of $d\approx\lambda$ and an asymmetric distribution of the reflective intensity (Fig. 1E). Since the incident angle is small, the diffraction is extremely surface sensitive. The depression metasurface can also obtain enhanced and diminished reflections in the same optical configuration. They are strong coherent plasmonic responses and resemble the asymmetric spectral feature of Fano resonances. Furthermore, there is a fundamental difference between the conventional non-plasmonic and plasmonic gratings that the plasmonic phenomenon occurs only for the TM(transverse magnetic) polarized light whose incident and grating vector are in the same plane and the magnetic field of the incidence is perpendicular to this plane (Fig. 1C). The strength of the plasmonic enhancement for a given wavelength is directly related to the spacing of the nanostructures. Based on the relation, we can establish our first encoding guideline that a given wavelength of light is represented by nanostructure arrays with a specific spacing. In general, we can define $\lambda=n_{eff}d$, where $d$ is the spacing and $n$ is effective refractive index. Factors such as surrounding medium, structure profile and fabrication accuracy can affect the effective refractive index. Nevertheless, it is a good working assumption that $n_{eff}\approx1$ for the air and metal interface in the encoded metasurfaces.

\paragraph*{}
Surface plasmon resonances are sensitive to their environment in the vicinity. The refractive-index sensitivity is the basis of plasmonic detection \cite{Anker2008}. If a metasurface is encoded with one-to-one correspondence of a given wavelength and uniformly contacts a medium, then detecting the change of the medium (refractive index) is equivalent to find the location where there are strong coherent plasmonic resonances. The change of the refractive index and its corresponding plasmonic structures can be written as $\Delta d=\lambda(n_{2}-n_{1})/n_{1}n_{2}$, where $n_{1}$ and $n_{2}$ are initial and changed refractive indices. Using the design (Fig. 1C) as a start point, we integrated detection and display functions into an encoded metasurface. We first demonstrate an approach to determine its sensitivity of the refractive index using the setup as illustrated in Fig. 2A. Fig. 2B shows experiment results using NaCl solutions of various concentrations (5\%, 10\%, 15\% and 20\% in deionized water). Responses of plasmonic resonances manifest the transmission efficiency as bright squares (Fig. 2B). The coherent coupling of the nanostructures in the Bloch mode requires matched phases. The phase delay in the medium meets the matching condition, where the incident light wavelength is approximately the value of \textit{n}(refractive index) times of the nanostructure periodicity ($\lambda_{0}\approx nd$). An average location shift, equivalent to 4 nm spectral shift, was observed using an 808 nm laser diode source. 

To determine the feasibility for practical biosensing applications, we verified the graphic pattern changes for detecting bindings of protein monolayers. Figs. 2C and 2D show two end-point measurements to detect the protein capturing functionality. As a proof-of-concept, high-affinity protein A/G and IgG were used in our experiments for sensing characterization \cite{Yanik2010a,Coskun2014,Xiao2016}. First, the gold-coated sensor surface was functionalized by protein A/G through physisorption. Capturing a single monolayer of bovine IgG antibody was revealed in both normal incidence using an 808 laser diode source (Fig. 2C) and grazing incidence mode using a 650 nm laser (Fig. 2D). In visible range (650 nm), the result is easily discernible by human eyes. Since the sensing process is presented as 2D graphics, digital image processing based on computer algorithms is helpful to analyze pictures and thus improve pictorial information for better human interpretation. The comparison image in Fig. 2C is an example, where the before and after photographs were combined together through a series of image processing. The red(before) and blue(after) squares indicate the positions of the corresponding sensing responses. The change of the position gives information of the biomolecular interaction. Most importantly, the sensing results can be further quantified by using different algorithms of image processing. For instance, the comparison image of Fig. 2D employed an alternative method to display the detection result. We decomposed pictures into separate RGB grayscale layers, then binarized one of them at a grayscale threshold (Fig. S4). After the process, it is surprisingly efficient to pinpoint the strongest spectral response of the nanostructure arrays within the periodic difference below 2 nm. This method simplifies the setting of plasmonic sensing. The conventional spectrum responses are encoded into the metasurface. It is also possible to improve the sensitivity by measuring intensity changes of each image pixel or using high-performance cameras with a wide dynamic range. Moreover, video cameras provide an option of recording the graphic changes to study kinetics (Movie S1). 

\paragraph*{}
The sensing application utilizes monochromatic light to establish a spectral relation to the encoded metasurface. Illuminated by a broadband/white light source, the encoded metasurfaces revealed another interesting phenomenon, which is a broadband optical response. Photographs captured by a digital camera as configured in Fig. 2A show the phenomenon for normal and grazing incidence modes. In the visible range, a single wavelength light corresponds to a spectral color. In the grazing incidence, the encoded plasmonic metasurface demonstrated the ability to distinguish a single wavelength light at a relatively narrow region (Fig. 3D), thus allowing us to define a one-to-one relationship for a full set of spectrum colors. Although the strongest transmission is dominated by the air and metal interface, there are interference in the normal incidence mode from two interfaces, resulting in enhanced and diminished transmission (Fig. 1E). These colors are more or less a combination of light with multiple wavelengths (Fig. 3C) and are affected by the supporting material. 

Colors in nature are not spectrally pure but mixed with different frequencies and intensities. Although the spectral responses of the periodic nanostructures are confined to a narrow spectrum, these responses can be mixed at distances far away from the order of the nanostructure size. The total intensity of light reflected or transmitted from a 2D surface is generally proportional to the area size. Being similar to altering the size of the area, varying the nanostructure length changes the surface coverage and thus provides an extra control of the incident light. It leads to our second encoding guideline that the relative intensity of corresponding light is defined by nanostructure length. Here, we use the red, green, and blue (RGB) model to demonstrate this encoding concept that produces a full-color image. Although the RGB model has narrower color gamut than the combination of full spectral colors, it is a simple and popular method to represent device-independent colors \cite{Hunt2004}. Three different periodic parameters were selected to represent three primary colors. The relative intensity of each primary color is tuned by the nanostructure length, which is proportional to the RGB values ranging from 0 to 255. The design is sketched in Fig. 3A. The scanning electron microscopy (SEM) images in Fig. 3B shows the details of the structural arrangement. The color-marked areas represent the RGB color pixels, which were encoded to display a mixed color according to its RGB value. The white dash square represents an RGB color pixel. The size of the RGB pixel is $1\times1$ $\mu$m and the width of a single pixel is 333 nm. Hence, a 333nm full-length nanostructure is defined as an RGB value of 255. Fig. 3E shows a photograph encoded with color images, which followed the described design rule. A tungsten-halogen incandescent lamp was used as the white light source. The periodic spacings are slightly different in each column of the pictures to demonstrate its hue adjustment.  

It should be noted that color models are based on human color perception and different colors can be produced from a combination of colors. It is not necessary to use a narrow spectral band for each primary color. Therefore, the same coding strategy can be expanded to non-plasmonic nanostructures for colorful images. Fig. 3F shows an image only using patterned electron beam resist (PMMA). It consists of a 60 nm PMMA layer spin-coated on a silicon substrate. The image encoded using the same RGB method was exposed and developed by electron beam lithography. Because of a relatively wide spectral band for each color, RGB graphic images using dielectric materials or non-plasmonic materials appears less color saturation than those coating with plasmonic metals. However, it is necessary to map the colors for different materials. On the contrary, the responses of encoded plasmonic metasurfaces are independent of the supporting materials. The established relation between the structure and color information can be readily applied to different material surfaces to achieve much the same color effect.  

\paragraph*{}
Metasurface encoding is a process that converts spectral information into a structural format based on a predefined structure and spectrum relationship. The potential of the encoded metasurfaces is not limited to spectral manipulation. For an incident light with a given wavelength, the distribution of its corresponding nanostructures can be arranged in a certain pattern, which can simultaneously shape light spectrally and spatially. Consider distances far from the periodically arranged nanostructures. We treat one nanostructure or several nanostructures in a periodic array as a point source. Interference from transmitted (Fig. S5) or reflected light from different positions can create spatial diffraction patterns. Computer-generated hologram(CGH) is a well-suited example, which is in essence a representation of optical diffraction patterns that are calculated from mathematical methods. We adapted our color image method and designed coding schemes to calculate the surface distribution of periodic nanostructures for spatial diffraction patterns. 

We established our third encoding guideline that the spatial diffraction is defined by the 2D distribution of plasmonic nanostructures, which only corresponds to a given wavelength. Spectral selectivity in the grazing incidence mode allows us to encode not only one but multiple images in a single hologram. The encoding method is analogous to the one generating the RGB images. Multiple independent wavelengths are assigned to different channels, which have their own periodic arrangements. A periodic and small pixelated area is treated as a point source, which can be occupied by several independent channels. Our computational method is based on the point source algorithm \cite{Rogers1950,Waters1966}. The multichannel approach is straightforward as illustrated in Fig. 4A, where each channel responds to one light wavelength. A SEM image of the fabricated multichannel hologram shows a phase binary CGH with 500 nm width and 2 $\mu$m period for each channel (Fig. 4B). The channel 1, 2, 3, and 4 have the corresponding wavelength of $\lambda_{1}$ = 650 nm, $\lambda_{2}$ = 532 nm, $\lambda_{3}$ = 450 nm, and $\lambda_{4}$ = 808 nm, respectively. The size of original binary images is $100\times100$ pixels. Each channel of the CGH has $600\times600$ pixels. All images (Fig. 4C) are reconstructed at the distance of 5 mm above the same surface illuminated separately using solid-state laser diodes. Every reconstructed image is in excellent agreement with its design and does not show interference from other channels. It is possible to further increase the number of the channels in one hologram using the same 4-channel design. Moreover, the design can implement simultaneous modulation of the amplitude and the phase. The modulation is revealed in a SEM image (Fig. 4E) where the phases are modulated by the distribution of nanostructures and their amplitudes are adjusted by the length. Fig. 4D is the reconstructed image. The grayscale of the original image was calculated to be proportional to the length of the corresponding nanostructures. Our CGH method serves as a demonstration of the encoding concept. More advanced algorithms or alternative arrangements of the nanostructures are certain to achieve faster calculation and better holographic images.        

\paragraph*{}
Our proposed metasurface encoding is based on a simple nanostructured surface relief. We found that there are great advantages in smoothing nanostructured plasmonic metasurfaces for spectral and spatial manipulation of light. The encoding methods involving the design of nanostructure size and distribution can be adapted to other nanostructure designs if the influence of supporting materials are carefully considered. The demonstrated applications still left room for new and improved functionalities such as combining sensing with color images or holograms, shrinking image pixels with a high refractive-index coating, creating true color 3D holograms and so on. More applications of the nanostructured metasurfaces, which benefit from the simple and flexible nanofabrication methods, are likely to be realized in a viable manner.     

\section*{Methods}
\paragraph*{High-throughput fabrication of nanostructured metasurfaces.}
We have developed a high-throughput nanofabrication technique for the nanostructured thin films. The technique adopted some procedures from nanoimprint and soft lithography to produce wafer-scale nanostructured surfaces, which transfers nanostructure patterns from a reusable master template to a variety of substrates via a stamp. The major steps of our fabrication technique are shown in Fig. S1. The master templates with nanostructure patterns can be produced by well-developed techniques, such as electron beam lithography (EBL), focus ion beam (FIB), interference lithography (IL), etc. We used EBL to create the nanostructure patterns on silicon substrates. To achieve smooth surfaces and a continuous metal coverage, we introduced a smoothing process. The surfaces of elevation nanostructures were spin-coated with a thin (10 nm) polymethyl methacrylate (PMMA) layer. For a depression type of nanostructures, a reflow process was used to smooth the PMMA resists that the developed resist is heated to its transition temperature 130$^{\circ}$C for 2 minutes. The smoothing process is a critical step to obtain ultrasmooth nanostructured surfaces and thus achieve coherent plasmonic couplings. After smoothing, the samples are ready for metal thin film deposition (50  nm Ag or Au) or are used as a master for high-throughput nanostamping fabrication. A transparent elastomer (polydimethylsiloxane, PDMS) stamp is made by casting PDMS on the master to duplicate the nanoscale features. Multiple PDMS stamps can be replicated from a single master, and the master and the stamp are reusable for the fabrication. 

\section*{Acknowledgments}
This work is supported by the NSF-CREST Grant number HRD 1547771.

\newpage
\printbibliography

\clearpage
\includegraphics[scale=0.5]{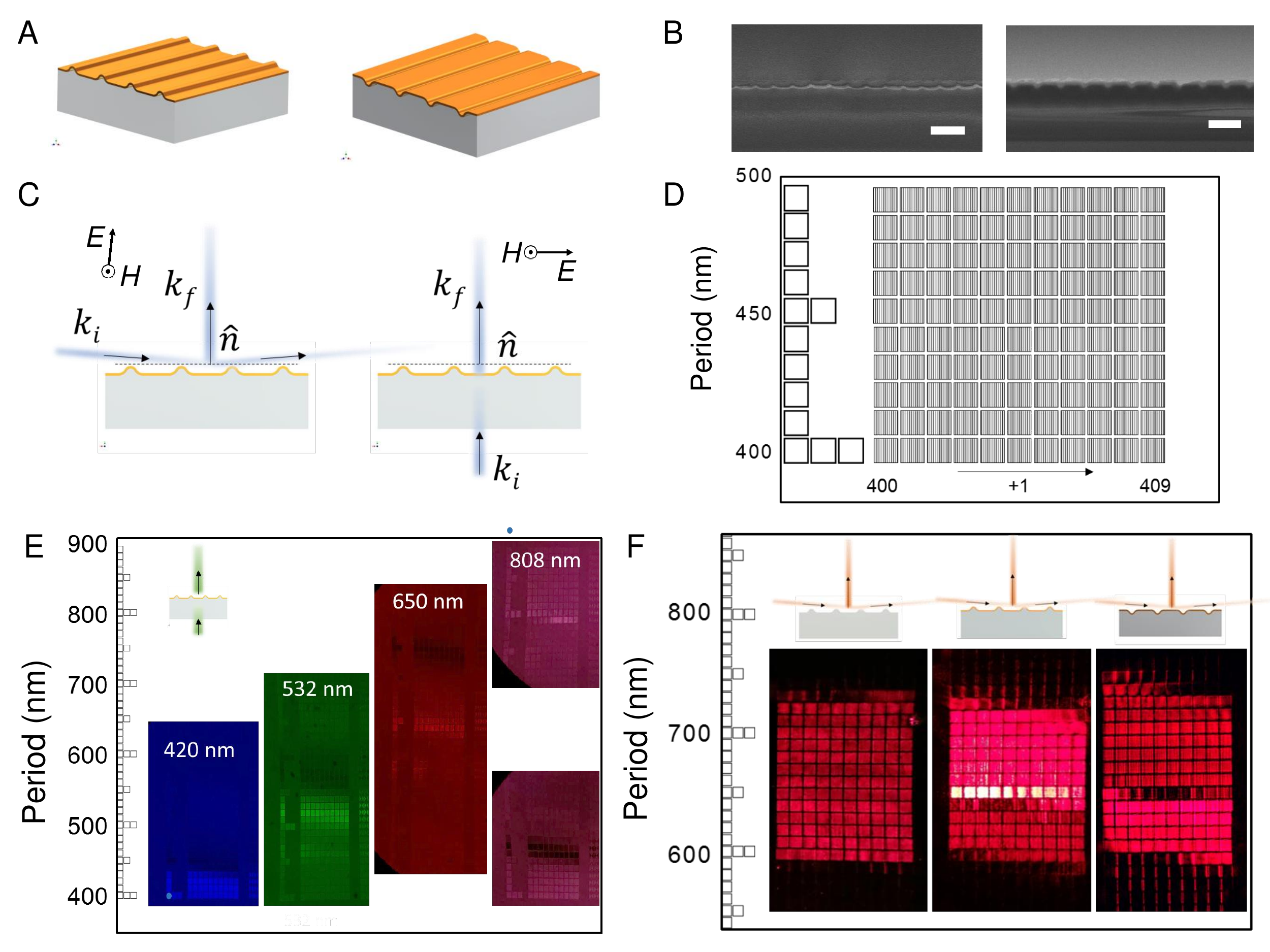}

\noindent {\bf Fig. 1. Nanostructured metasurfaces.} (A) Illustrations of the elevation and depression surface reliefs. (B) Side-view SEM images of the elevation and depression metasurfaces. The scale bar is 1 $\mu$m. (C) Schematics of two incidence operation modes. (D) A layout design of the metasurface with nanostructure arrays. Every square represents one periodic nanostructure array. The increment of the period is 1 nm from left to right and 10 nm from bottom to top. (E) Photographs of the metasurface illuminated using laser diode sources with the wavelengths of 420, 532, 650 and 808 nm. The inset shows the incidence direction. (F) Photographs of the metasurface under a grazing angle illumination using a 650 nm laser diode source. From left to right, an elevation metasurface made of SU-8 epoxy resist on a glass substrate without a metal coating, with a 40 nm silver coating and a depression metasurface with a 40 nm silver coating. 

\includegraphics[scale=0.5]{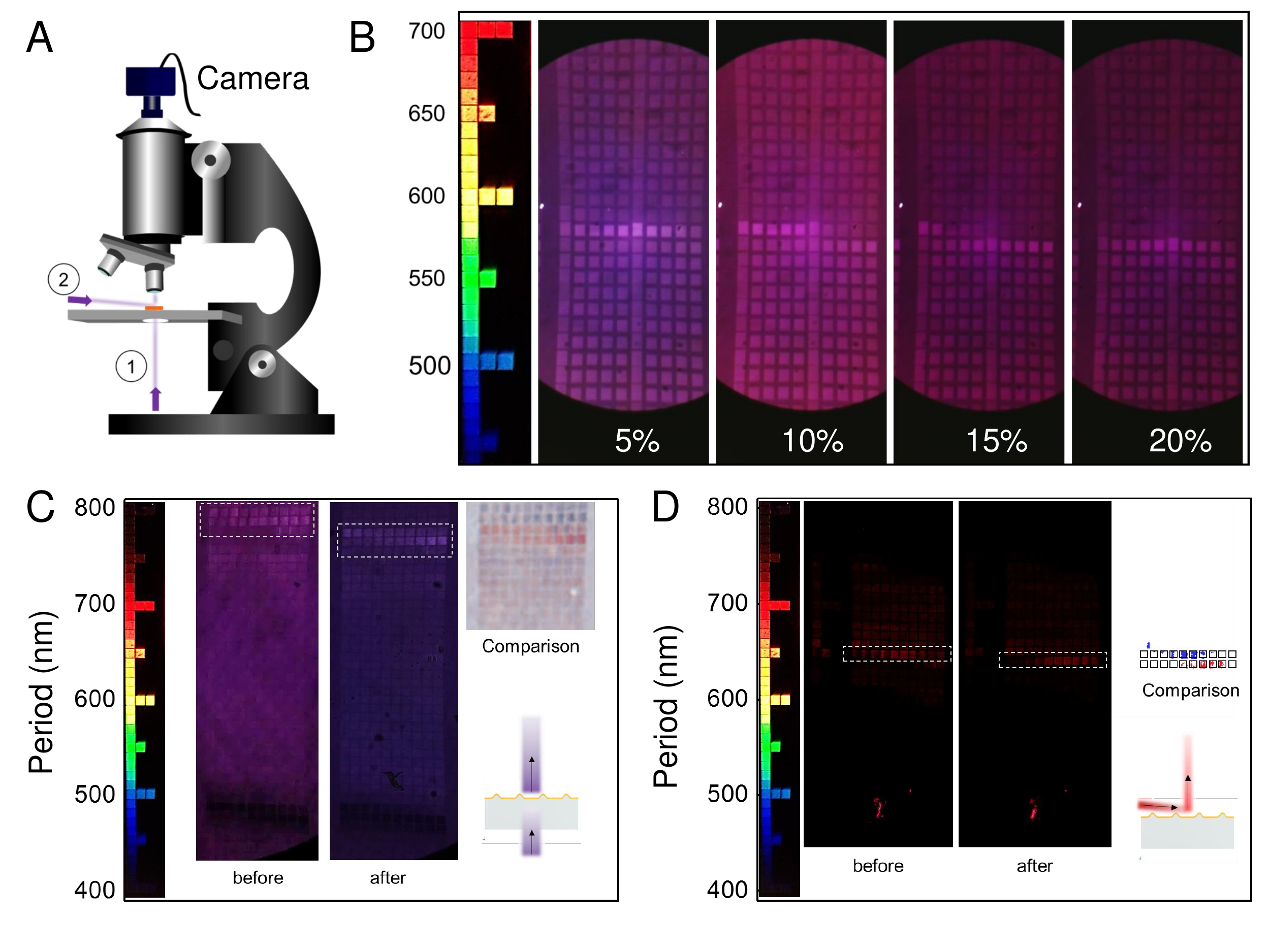}

\noindent {\bf Fig. 2. Encoding metasurfaces for chemical and biosensing.} (A) Schematic diagram of the sensing setup under two incidence operation modes: 1. normal incidence and 2. grazing incidence. (B) Photographs of the sensor surface in NaCl solutions of various concentrations, illuminated using an 808 nm laser diode under normal incidence. The concentrations of the NaCl solutions were 5\%, 10\%, 15\% and 20\% in deionized water, corresponding to refractive indices of 1.3418, 1.3505, 1.3594, and 1.3684, respectively. (C) Photographs of the sensor surface immobilized with protein A/G before and after capturing a monolayer bovine IgG antibody, illuminated using an 808 nm laser diode under normal incidence. The comparison image combines the before (blue) and after(red) images into one. (D) Photographs of the sensor surface immobilized with protein A/G  before and after capturing a monolayer bovine IgG antibody, illuminated using a 650 nm laser diode at a grazing angle. The comparison image combines the binarized before(blue) and after(red) images.

\includegraphics[scale=0.5]{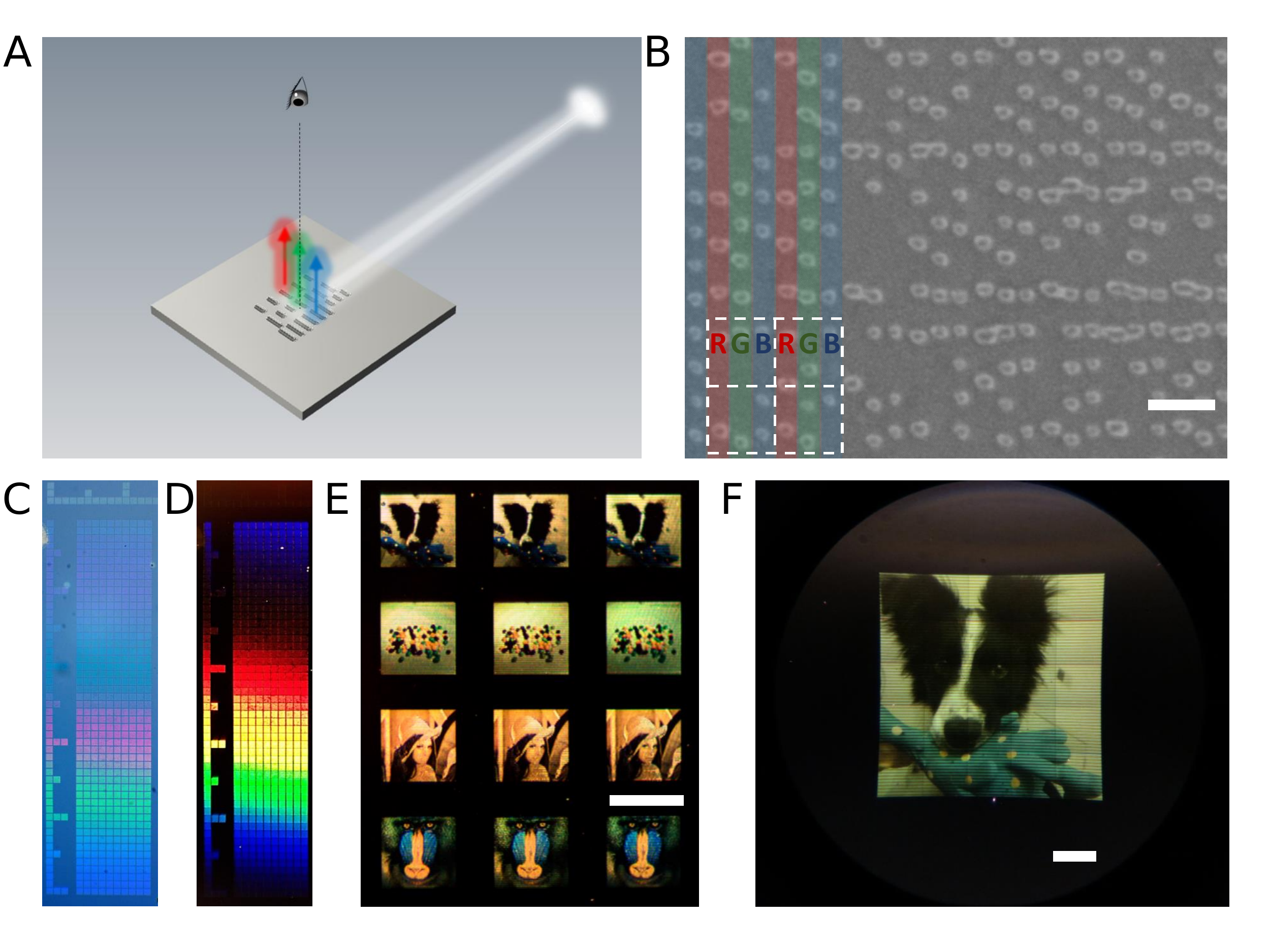}

\noindent {\bf Fig. 3. Encoding metasurfaces for colorful images.} (A) Illustration of displaying a colorful image from an encoded metasurface. (B) SEM image of the encoded metasurface. The white dash square represents an RGB color pixel, which has the size of $1\times1$ $\mu$m. The width of a single color channel in an RGB pixel is 333 nm. The scale bar is 1 $\mu$m. (C and D) Photographs of the elevation metasurface with nanostructure arrays, which have the periods from 400 to 899 nm (C) under normal incidence of a white light source and (D) under grazing incidence of a white light source. (E) Photograph of the silver-coated metasurface encoded with color images. From left to right column, the periodic spacings corresponding to the colors of (red ,green, blue) are (670, 510, 440), (680, 535, 440) and (690, 510, 450) (nm). (F) Photograph of the metasurface made of PMMA resist on a Si substrate and encoded with a color image under grazing incidence of a white light source. The scale bar of (E) and (F) is 100 $\mu$m.

\includegraphics[page=1,width=0.5\columnwidth]{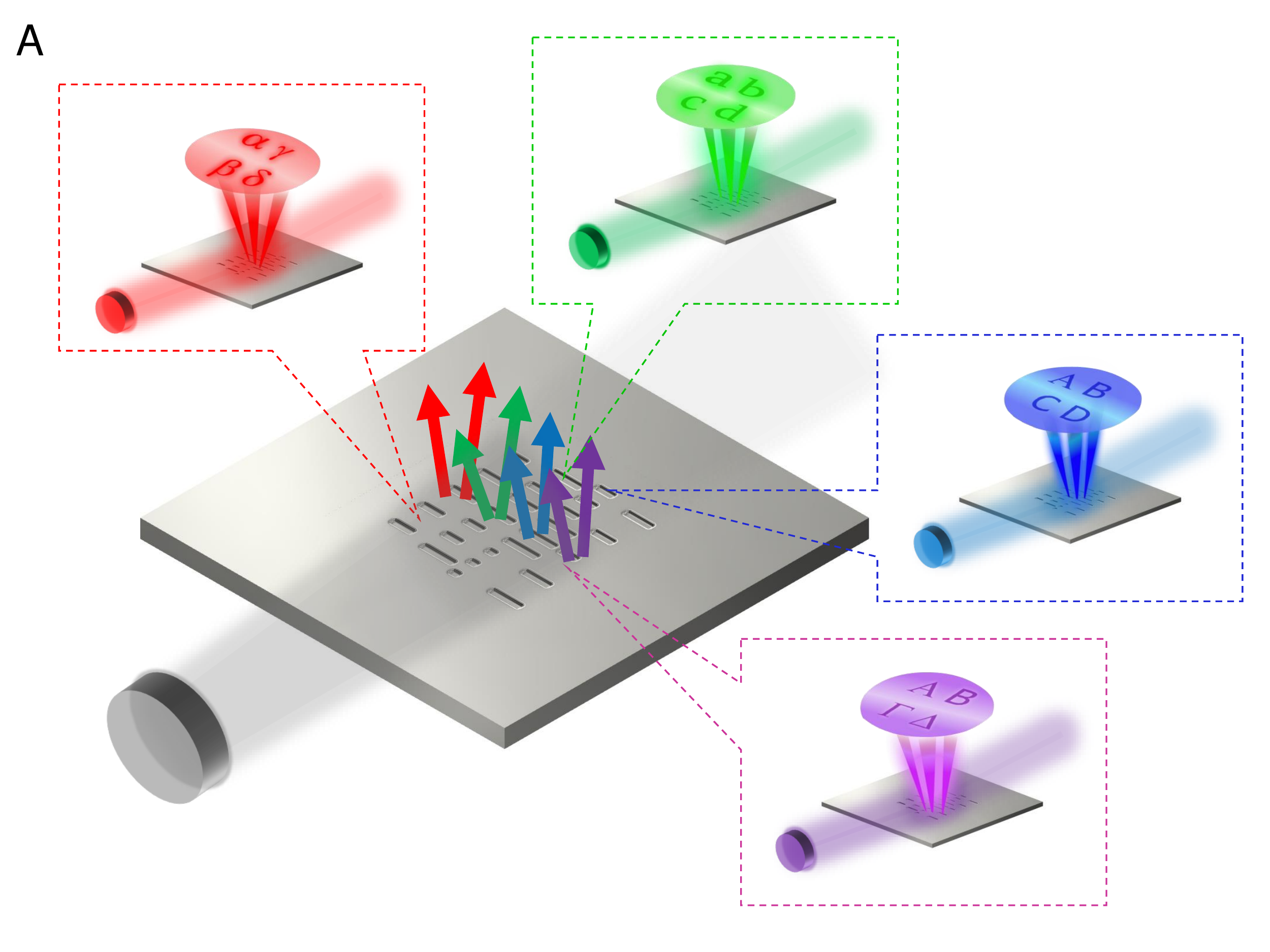}
\includegraphics[page=2,width=0.5\columnwidth]{Figure4}

\noindent {\bf Fig. 4. Encoding metasurfaces for holograms.} (A) Illustration of a multichannel hologram. (B) SEM image of the hologram encoded with four channels. The designated wavelengths for the four channels: $\lambda_{1}$ = 650 nm, $\lambda_{2}$ = 532 nm, $\lambda_{3}$ = 450 nm, and $\lambda_{4}$ = 808 nm. The width and the period of a single channel are 500 nm and 2 $\mu$m. The scale bar is 1 $\mu$m. (C) Four reconstructed holographic images illuminated separately using laser diode sources with the wavelengths of $\lambda_{1}$, $\lambda_{2}$, $\lambda_{3}$, and $\lambda_{4}$. (D) SEM image of the hologram encoded with amplitude and phase modulation. The amplitude is modulated by the length of the nanostructures. It is a single channel hologram in which both the width and the period of the channel are 500 nm. (E) Reconstructed holographic image from the encoded metasurface with amplitude and phase modulation. The inset is the original digital image, which has the size of $100\times100$ pixels.

\newpage
\begin{center}
\section*{Supplementary Materials for\\
    Encoding Multifunctional Nanostructured Metasurfaces}

    Bo Xiao, Christian G. Carvajal, Sangeeta Rout
    \\
    \normalsize{Center for Materials Research, Norfolk State University,}\\
    \normalsize{700 Park Ave, Norfolk, Virginia 23504, USA}\\
\end{center}

\subsection*{Materials and Methods}
\subsubsection*{i. High-throughput fabrication method}

\begin{center}
 \centering
 \includegraphics[width=160mm]{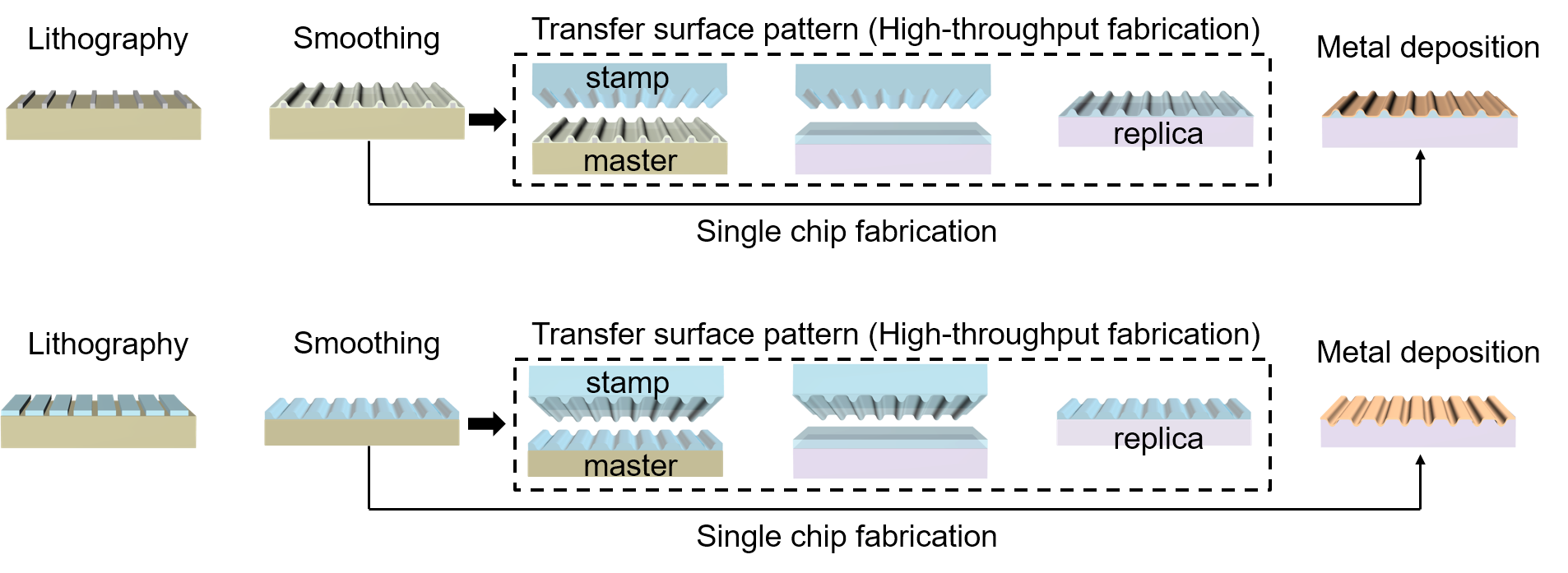}
\end{center}
Fig. S1 Fabrication process flow diagrams of the nanostructured metasurfaces. a. Surface elevation structures. b. Surface depression structures.

\subsubsection*{ii. Plasmon resonance coupling}
The metasurfaces with the elevation and the depression surface reliefs exhibit different transmission responses under normal incidence of light. We content ourselves with a brief explanation and simply treat the whole structure as two separated parts (Fig S2). Consider plasmonic resonances on both sides of the metal thin films induced by a normally incident plane wave. The coherent sum of the transmitted field will be significantly affected if the structure is largely surrounded by its supporting medium. We can also make a reasonable prediction that the conventional grating profiles(sinusoidal, blaze angle) would not be suitable for our proposed encoding due to the dependence of the supporting materials (Fig. S2). In addition, the enhanced transmission is irrelevant whether the incident light is illuminated from the air side or the substrate side. It implies that the curvature of the nanostructures plays an important role in plasmonic radiation and the dominant contribution to the transmission enhancement comes from the convex surface. The detailed discussion is beyond the scope of this study and will be the object of future work.
\begin{center}
 \centering
 \includegraphics[width=100mm]{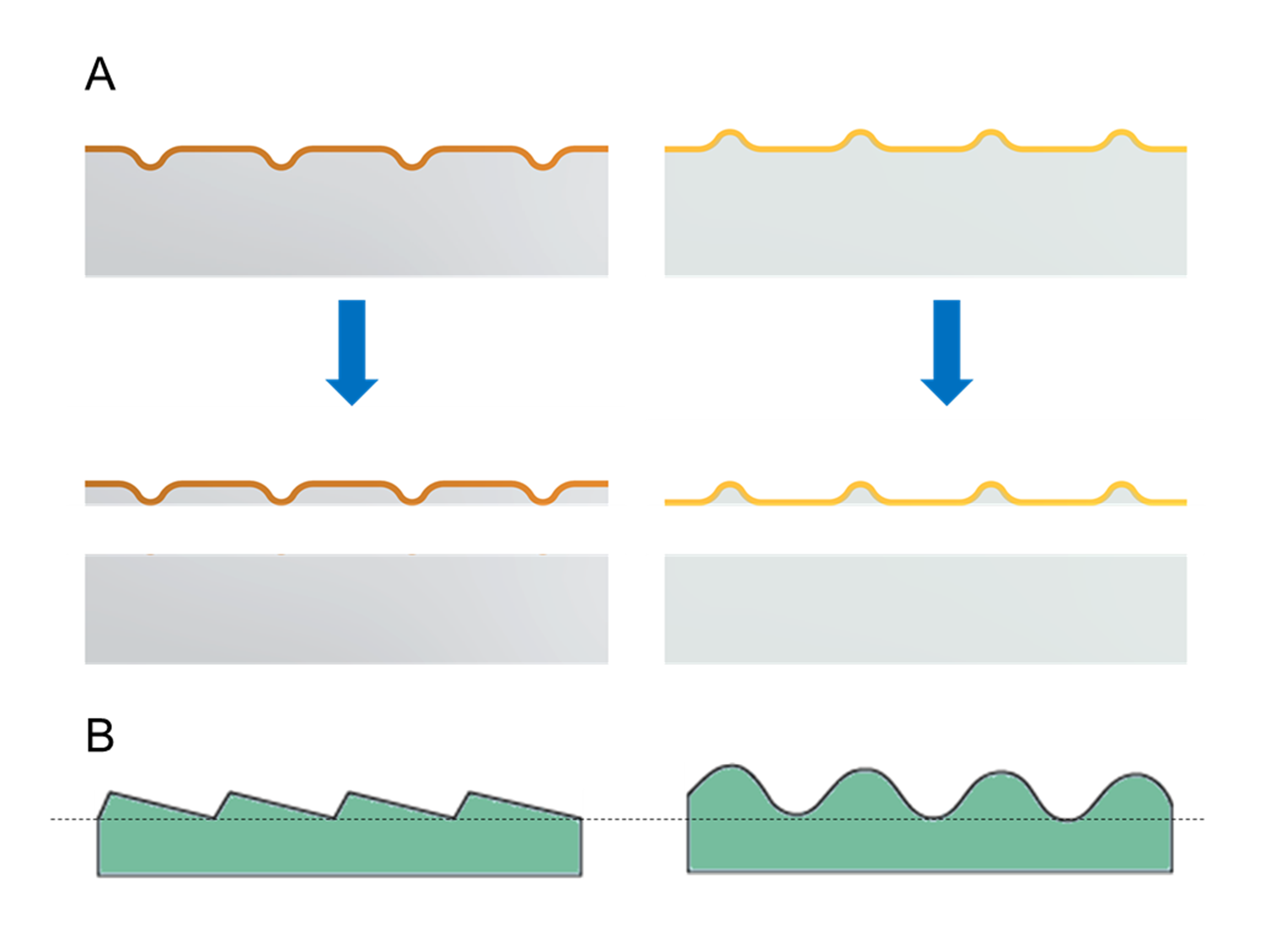}
\end{center}
Fig. S2 a. A simple model of elevation and depression surface relieves. b. The profiles of the conventional sinusoidal and blaze angle gratings.

In one- and two-dimensional plasmonic nanostructures, the optical responses can produce collective plasmon resonances through dipole coupling. In the optical spectra, the intensity and the position of plasmonic resonances manifested as sharp spectral peaks depend on the distribution of the nanostructures and the dielectric function of the constituent materials. The nanostructures are small compared to the wavelength (even after smoothing). They are therefore treated in a point dipole approximation. As the continuous metal thin films can suppress the interference from the dielectric interface, the coupled-dipole approximation can be utilized, treating the periodic nanostructures with localized surface plasmon resonances as the simplest case of a one-dimensional array of interacting atomic dipoles on the boundary of an infinite half-space of vacuum ($\epsilon=1$). The cooperative behavior can significantly modify the optical response, which is analogous to the modified behavior of a single quantum emitter inside a cavity \cite{Eschner2001,Bettles2016}. In Fig. S3, we plot the decay rate of a single atomic dipole inside a cavity mirror that sheds light on the spacing dependence of the dipole-dipole interactions. The half-decay rate \cite{Milonni1973} is $$\gamma = \frac{3\pi \gamma _{0}}{k_{0}a}\left [ \frac{1}{2} +\sum_{n=1}^{[k_{0}a/\pi]}\left ( 1- \frac{n^{^{2}}\pi^{2} }{k^{2}a^{2}}cos^{2}\left ( \frac{n\pi }{2} \right )\right )\right ]$$, where $k_{0}=\omega_{0}/c$, $[k_{0}a/\pi]$ is the greatest integer part of $k_{0}a/\pi$, \textit{n} is the cavity mode index, \textit{a} is the mirror spacing, and $\gamma_{0}$ is the half-decay rate of a single atom in free space.
\begin{center}
 \includegraphics[width=100mm]{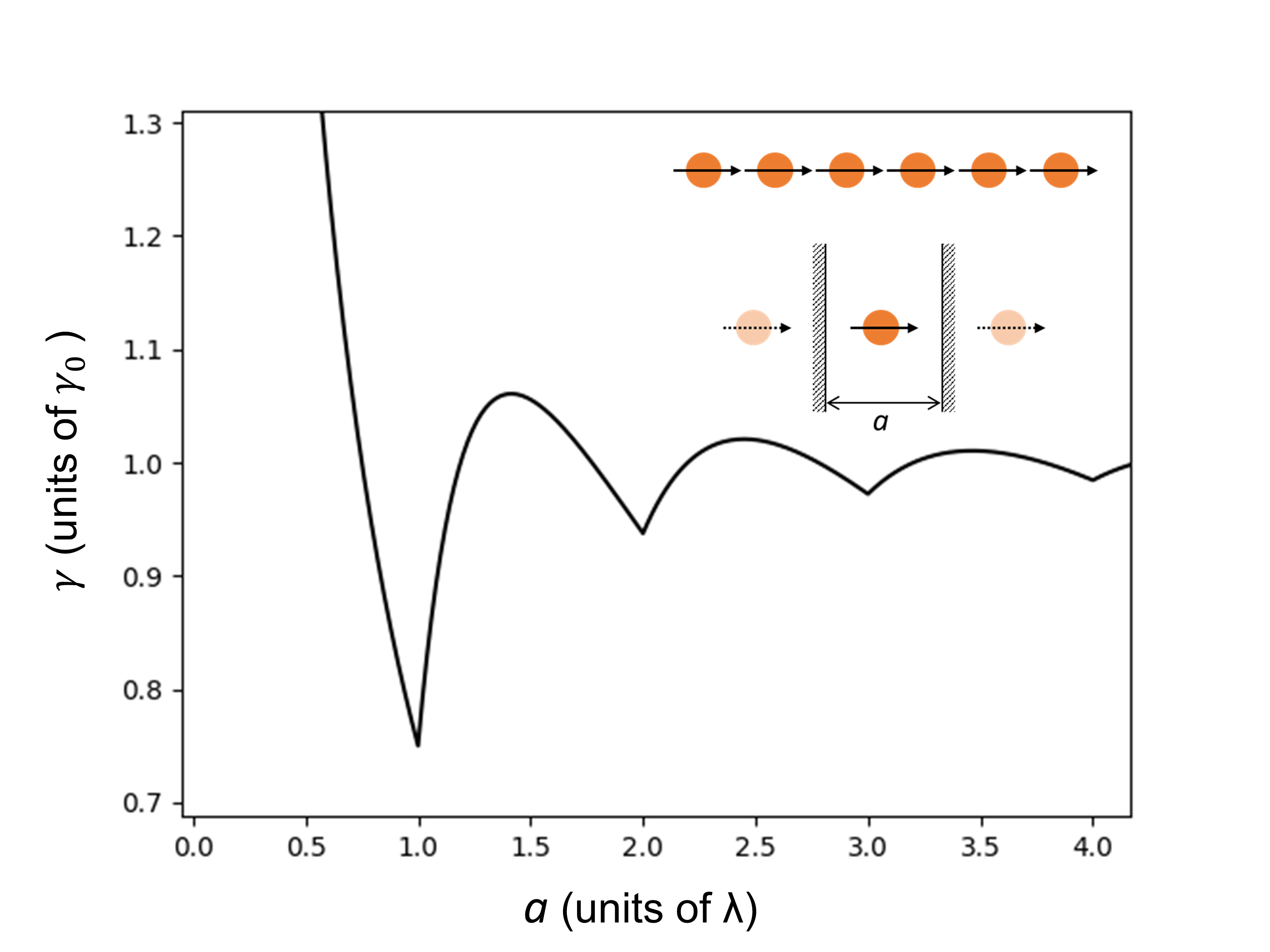}
\end{center}
Fig. S3 Half-decay rate of a single atomic dipole inside a cavity mirror.   

\subsubsection*{iv. Image Capturing Setup and Image Processing}
A consumer-grade microscope was modified to capture all the images. The original light source of the microscope was replaced with laser diodes and a fixture was set aside for mounting grazing angle light sources. Color and near-infrared images were captured by a Raspberry Pi infrared camera module (Pi NoIR V2 IMX219 8-megapixel sensor), which was mounted on the eyepiece of the microscope. The camera has the same functions as the regular module but without an infrared filter. Blue and red filters leak infrared light. Under the infrared illumination, a mixture of signals from red and blue pixel sensors is interpreted as purple. All adjustments of contrast and brightness were applied to an entire image. Fig. S4 shows an image process to determine the peak position.    
\begin{center}
 \includegraphics[width=160mm]{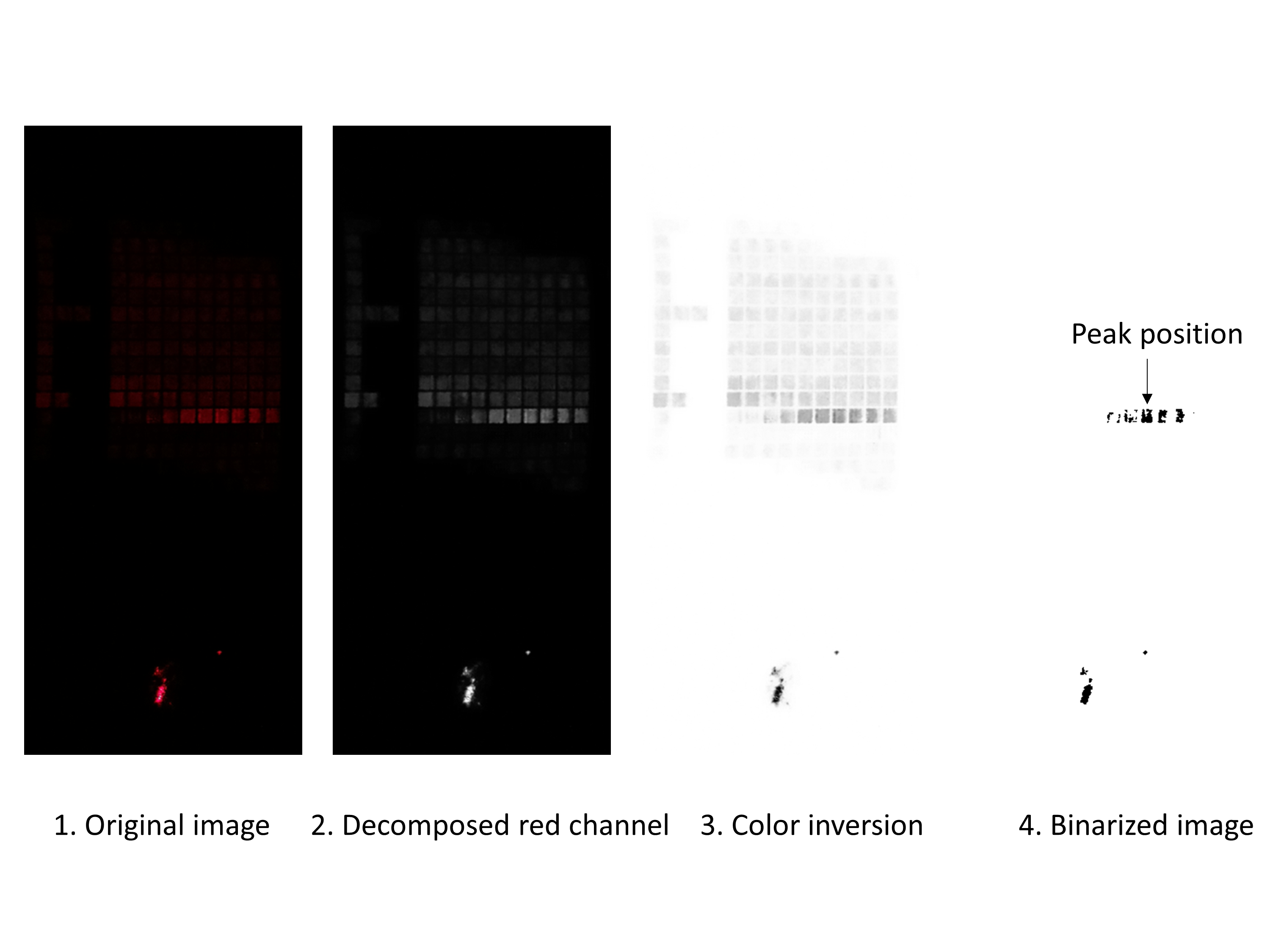}
\end{center}
Fig. S4 Image process steps to determine the peak position.

\subsubsection*{v. Kinetics and affinity measurements using encoded metasurfaces}
The encoded metasurfaces are capable of performing the kinetic analysis. Movie S1 is a video clip of the formation of a thiol-SAM (self-assembled monolayer) on the gold-coated metasurface. To obtain a large and uniform light distribution, a rotating diffuser was placed in front of an 808 nm laser diode. It's the cause of the background flickering in the video. 

The SAM was prepared using two thiols, 2.5 mM 11-mercaptoundecanoic acid (MUA) and 7.5mM 1-decanethiol in ethanol solution. Movie S1 is a 10x fast forward video which has a timestamp on the top of the screen. The sample was immersed in a phosphate-buffered saline (PBS) solution at the recording start. The SAM solution was injected at the timestamp of 00:53 (Movie S1). 

\subsubsection*{vi. Transmission Holograms}
Fig. S5 shows two binary transmission holograms designed for the wavelengths of 808 and 532 nm. Nanostructure arrays with two different periodic spacings can define the transmission efficiency as 0 and 1 (lowest and highest transmission). It is a straightforward method to determine the structure parameters by locating the enhanced/diminished transmission in Figure 1e. The nanostructure periods of (470 nm, 796 nm) were used to define the transmission (0, 1) for the wavelength of 808 nm, and the (555 nm, 522nm) was for the wavelength of 532 nm.
\begin{center}
 \includegraphics[width=160mm]{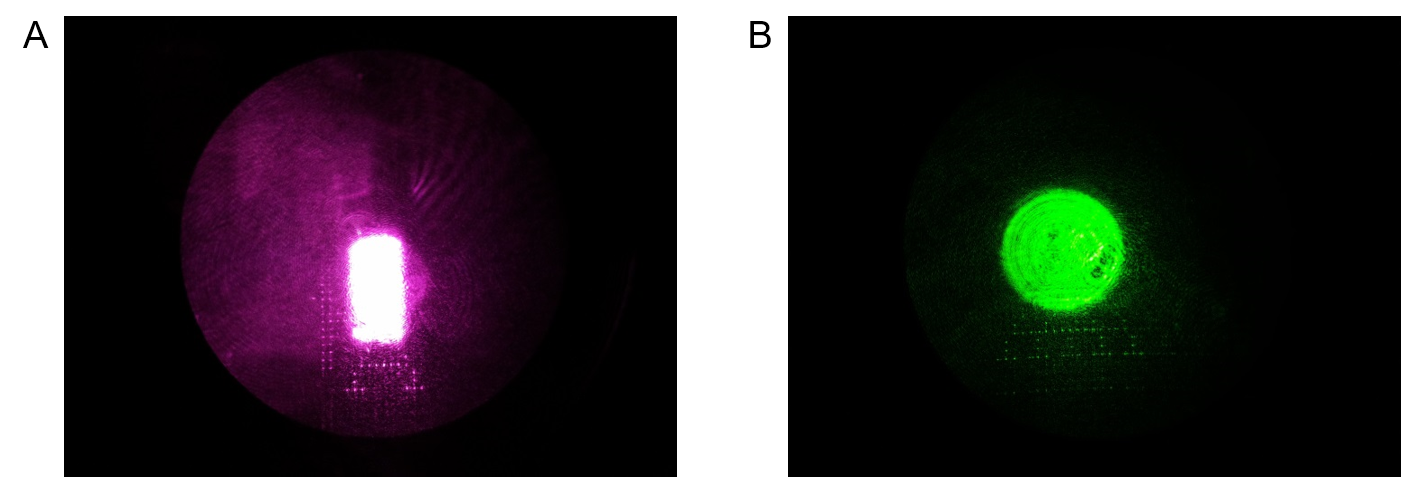}
\end{center}
Fig. S5 Reconstruction of the binary transmission holograms for the wavelengths of (a) 808 and (b) 532 nm.

\end{document}